 \documentclass[prb,aps,showpacs,twocolumn,a4paper]{revtex4}
 \usepackage{graphicx}

\begin{document}

  \title{Electronic structure study by means of X-ray spectroscopy
and theoretical calculations of the ``ferric star'' single molecule
magnet}
\author{A. F. Tak{\'a}cs}
\email{altakacs@uni-osnabrueck.de} 
\affiliation{Universit\"at Osnabr\"uck -- Fachbereich Physik, 
D-49069 Osnabr\"uck, Germany}
\author{M. Neumann}
\affiliation{Universit\"at Osnabr\"uck -- Fachbereich Physik, 
D-49069 Osnabr\"uck, Germany}
\author{A.~V.~Postnikov}
\email{apostnik@uos.de} 
\affiliation{Institut f\"ur Festk\"orperforschung, 
Forschungszentrum J\"ulich, D-52425 J\"ulich, Germany 
{and} Russian Academy of Science -- Ural Division,
S. Kowalewskoj 18, 620219 Yekaterinburg, Russia}
\author{K. Kuepper}
\affiliation{Universit\"at Osnabr\"uck -- Fachbereich Physik,
D-49069 Osnabr\"uck {and} Forschungszentrum Rossendorf, Institut
f\"ur Ionenstrahlphysik und Materialforschung, Bautzner Landtrasse
128, 01328 Dresden, Germany}
\author{A. Scheurer}
\affiliation{Universit\"at Erlangen--N\"urnberg,
Insitut f\"ur Organische Chemie, D-91054 Erlangen, Germany}
\author{S. Sperner}
\affiliation{Universit\"at Erlangen--N\"urnberg,
Insitut f\"ur Organische Chemie, D-91054 Erlangen, Germany}
\author{R. W. Saalfrank}
\affiliation{Universit\"at Erlangen--N\"urnberg,
Insitut f\"ur Organische Chemie, D-91054 Erlangen, Germany}
\author{K. C. Prince}
\affiliation{Sincrotrone Trieste, Strada Statale 14, km 163.5, in
Area Science Park, I-34012 Basovizza (Trieste), Italy and
INFM-TASC, Laboratorio ELETTRA, I-34012 Basovizza (Trieste), Italy}
\date{November 21, 2005}
\begin{abstract}
The electronic structure of the single molecule magnet system
\{$M$[Fe(L$^1$)$_2$]$_3$\}$\cdot$4CHCl$_3$ ($M$ = Fe, Cr ; L$^1$ =
CH$_3$N(CH$_2$CH$_2$O)$_2$$^{2-}$) has been studied using X-ray
photoelectron spectroscopy, X-ray absorption spectroscopy, soft
X-ray emission spectroscopy, as well as theoretical density
functional-based methods. There is good agreement between
theoretical calculations and experimental data. The valence band
mainly consists of three bands between 2 eV and 30 eV. Both theory
and experiments show that the top of the valence band is dominated
by the hybridization between Fe $3d$ and O $2p$ bands. From the
shape of the Fe $2p$ spectra it is argued that Fe in the molecule
is most likely in the 2+ charge state. Its neighboring atoms (O,
N) exhibit a magnetic polarisation yielding effective spin $S$=5/2
per iron atom, giving a high spin state molecule with a total
$S$=5 effective spin for the case of $M$ = Fe.
\end{abstract}
 \maketitle
\section[#1]{Introduction}
\label{sec:intro}
In the quest for molecular magnets which might be useful for
practical applications such as ultradense magnetic storage
\cite{PhilTransA357-1762}, quantum computing \cite{Nat410-789}, or
other interesting devices \cite{ElChInt11-34}, one finds at times
molecules which combine structural beauty, promising chemistry for
further development, non-trivial physics and practically
interesting properties (intramolecular exchange interactions,
magnetic anisotropy). Such new molecular materials often arise
from metal-organic synthesis, which is the main driving force in
studies of molecular magnets (see, e.g., the monograph by Kahn
\cite{Kahn-book}, or
Refs.~\cite{Mol_Magnets,MRSB25-21,JPCM16-R771} for recent
reviews). The present work is devoted to one such system, known as
``ferric star'', which is simple enough to allow an accurate study
of its electronic structure, yet far from trivial for what regards
its chemical manipulation (crystallization with different ligands)
and physical treatment (e.g., deposition on surfaces). The
synthesis of magnetic molecules with the formula
\{$M$[Fe(L$^1$)$_2$]$_3$\}$\cdot$4CHCl$_3$, where L$^1$ $\equiv$
CH$_3$N(CH$_2$CH$_2$O)$_2$$^{2-}$ is an organic ligand and
\emph{M} is Fe or Cr, has been described by Saalfrank \emph{et
al.} \cite{ChEurJ7-2765} (compounds {\bf 11} and {\bf 12} of this
paper). Previous basic magnetochemical investigations of the
``ferric star'' \cite{JACS121-5302} indicate that the three spins
$S$=5/2 of peripheral Fe ions at the star edges couple with an
antiparallel orientation to the central ion, yielding (for
\textit{M} = Fe) the magnetic ground state of $S$=5. In the
present work we attempt to clarify the electronic structure, as it
emerges from a combined spectroscopic study and first-principles
calculations.

\begin{figure*}[t]
\centerline{\includegraphics[width=0.9\textwidth,clip=true]{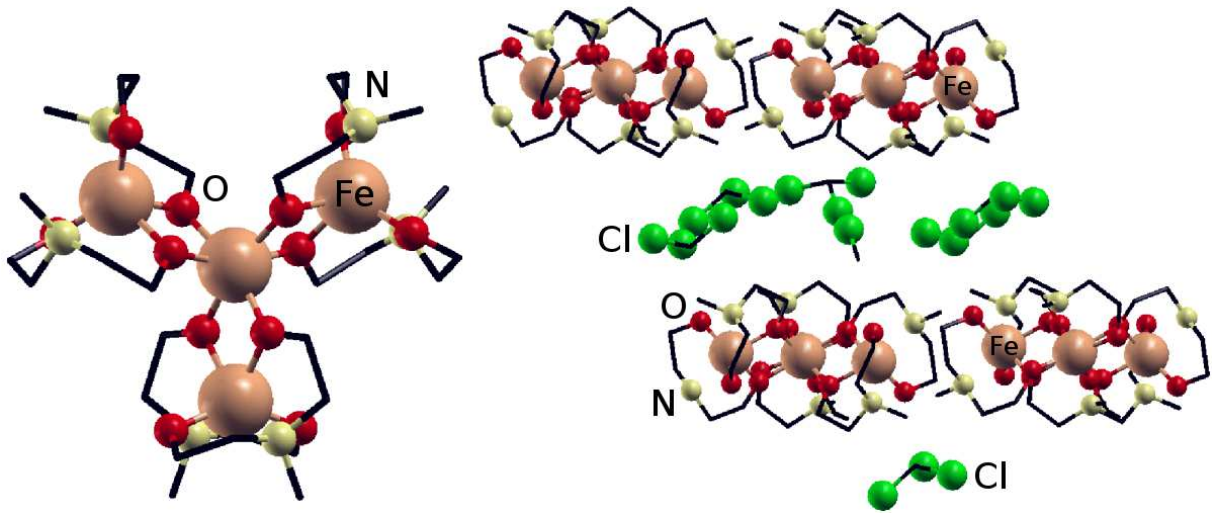}}
\caption{\label{fig:molec}
Single ``ferric star'' molecule in the top
view (left) and side view of the molecules packed in the crystal
structure with chloroform (right). Hydrogen atoms are omitted, and
carbon chains are shown as wireframe.}
\end{figure*}

Specifically, we probe core-level excitations of the ferric star
at the \textit{L} and other edges by several core spectroscopies,
namely near-edge X-ray absorption spectroscopy (NEXAS), resonant
inelastic X-ray scattering (RIXS) and X-ray photoemission
spectroscopy (XPS). The electronic structure calculations were
done from first principles within the density functional theory.
The spectra give insight into the distribution of energy-resolved
state densities within the valence band, which reveal the $M$ $3d$
-- O $2p$ hybridization, and into charge states of transition
metal (TM) atoms. The calculations give estimates of local
magnetic moments, spatial distribution of the spin density, and of
magnetic interaction parameters.

The paper is organized as follows. In Sec.~\ref{sec:struc} we
summarize present knowledge of the crystal structure and basic
magnetic properties. Sec.~\ref{sec:details} gives technical
details of the spectroscopy measurements, and of the
first-principles electronic structure calculations. The discussion
of results and comparison with the calculations is done in
Sec.~\ref{sec:results}.

\section[#2]{Structure and magnetic properties}
\label{sec:struc}
The samples for our present spectroscopy study, as well as the
crystallographic data used in electronic structure calculations,
originate from the the Institute of Organic Chemistry of the
University of Erlangen -- N\"urnberg \cite{ChEurJ7-2765}. The
precise chemical formula of the compound in question is
\{$M$[Fe(L$^1$)$_2$]$_3$\}$\cdot$4CHCl$_3$, where \emph{M} is Fe
or Cr in the present study. In the following we refer to these
systems as FeFe$_3$ and CrFe$_3$-stars. The structure of a single
molecule derived from crystal structure analysis can be seen in
the left panel of Fig.~\ref{fig:molec}, while the spatial packing
of the molecules crystallized from chloroform is shown in the
right panel. The central iron ion is linked via two $\mu_2$-alkoxo
bridges from each of the three terminal building blocks
[Fe(L$^1$)$_2$]$^{-}$. The peripheral iron centers are
octahedrally coordinated through two N--, $\mu_1$O--, and
$\mu_2$O--donors. All four iron ions are located in a plane, with
Fe--Fe--Fe angles of about
120$^{\circ}$. 
As known from the literature, similar types of ferric stars with
the FeFe$_3$-- framework can be synthesized using different
ligands like dipivaloylmethane (Hdpm) in combination with
methanolate or 1,1,1 -- tris(hydroxymethyl)ehane (H$_3$thme),
yielding respectively [Fe$_4$(OMe)$_6$(dpm)$_6$]
\cite{JACS121-5302} or [Fe$_4$(thme)$_2$$_6$]
\cite{AngewChem2004-43}.
 The chemistry of \{$M$[Fe(L$^1$)$_2$]$_3$\}$\cdot$4CHCl$_3$
 has certain similarities with that of
``ferric wheels'' \cite{AnChIE36-2482}, which have the same
ligands and a similar (nearly octahedral) coordination of the Fe
ions. A relation between the two systems is demonstrated by the
fact that adding two equivalents of iron (III) chloride to
[\emph{M}\{Fe(L$^1$)$_2$\}$_3$] produces the known ``ferric
wheel'' with a metallacrown structure as the final product
\cite{ChEurJ7-2765}. As in these latter compounds which we have
studied earlier \cite{EMRS-Fewheel,Bedlewo-Fewheel}, the spins
$S$=5/2 of Fe ions tend to couple antiparallel. Contrary to
``ferric wheels" with even numbers of Fe atoms, which have zero
spin in the ground state, in the ``ferric star'' the antiparallel
coupling of three outer Fe spins to the central one results in a
net ground-state spin $S=3\times 5/2 - 5/2 = 5$. For the magnetic
anisotropy, the effective Hamiltonian of which can be
conventionally written in the form ${\cal H}=DS_z^2 +
E(S_x^2-S_y^2)$ in terms of the cumulative spin of the molecule,
${\bm S}$the axial parameter $D$ is $-$0.20 cm$^{-1}$ ($-$0.29 K)
from the high field electron spin resonance experiments
\cite{JACS121-5302,ChemComm2000-725}. The absolute value of $E$
(the sign may depend on the definition) was predicted by Kortus to
be 0.064 K from a first-principle calculation \cite{Kortus_aniso},
and later estimated as 0.056 K in experiments by M\"uller \emph{et
al.} \cite{Mueller_aniso}.

It has been recently demonstrated that Fe in ``ferric star'' can
be substituted by other $3d$ ions. In the present work we study
the spectroscopy of Cr-doped ``stars''. The accuracy of the amount
and position of the substitution has not yet been unambiguously
established. For ab initio calculations we assumed the
substitution in the central position, in an otherwise unchanged
geometry. A more systematic theoretical analysis of the electronic
structure of ``ferric wheels'' with substitution by different ions
in different positions will be reported elsewhere
\cite{stars_theo}.

\section[#3]{Experimental and calculation details}
\label{sec:details}
The XPS measurements were performed using a PHI Model 5600ci
MultiTechnique system in the Dept. of Physics, University of
Osnabr\"uck. The Al$K_{\alpha}$ radiation was monochromatized by a
double focusing monochromator and the pressure during the
measurements was typically 10$^{-9}$ mbar. The total energy
resolution, as determined at the Fermi level of a gold foil, was
0.3--0.4 eV. The resolution of the electron energy analyzer was 80
meV. The samples were insulating, so charge neutrality on the
surface was achieved using a low energy electron flood gun. The
recording time of the presented spectra was less than 10 h. The
NEXAFS and RIXS measurements where carried out at the BACH
beamline of the synchrotron radiation facility in Trieste, using
the COMIXS spectrometer \cite{COMIXS,BACH}.

Normal X-ray emission spectroscopy was performed also at the ROSA
beamline at the BESSY II Synchrotron facility in Berlin, using the
undulator based beamline ID12-2, and the rotatable spectrometer
apparatus (ROSA) at BESSY II, Berlin. The excitation energies were
set to 750 eV for the Fe $L$ edge, to 420 eV for the N $K$ edge,
and to 600 eV for the O $K$ edge.  The overall resolution
(beamline plus spectrometer) was set to around 1 eV.

For X-ray photoemission spectroscopy, emitted electron energies
were calibrated to the C $1s$ line from CH$_2$--CH$_2$--O
\cite{Beamson}. For X-ray absorption the resolution was set to 0.3
eV and for emission to 0.7 eV. The emitted photon energies at the
iron edge were calibrated to the Fe metal $L_{\alpha}$ normal
emission maximum at 705 eV and at the oxygen $K$ edge to the
emission from MgO at 525 eV. The pressure during the measurements
was 10$^{-9}$ mbar.

The X-ray photoelectron spectra were recorded of Fe ($3s$, $2p$),
C $1s$, O $1s$, and N $1s$ core levels. The main peaks of the
Fe$3s$ spectra of the ``ferric star'' were fitted with a Voigt
function \cite{DS_func}, by constraining the Gaussian width to a
value of 0.6 eV. The background was simulated by a Tougaard
function \cite{Toug_func}, and subtracted from the spectra.

The electronic structure of the ``ferric stars'' was calculated
from first principles within the density functional theory, using
the calculation method and computer code {\sc Siesta}
\cite{PRB53-10441,JPCM14-2745,siesta}. The method uses
norm-conserving pseudopotentials in combination with atom-centered
strictly confined numerical basis functions
\cite{JPCM8-3859,PRB64-235111}. The basis set included
double-$\zeta$ functions with polarization orbitals added for Fe
and O. The treatment of exchange-correlation was done according to
the generalized gradient approximation after
Perdew--Burke--Ernzerhof \cite{PRL77-3865}.

The single molecular unit \{\emph{M}[Fe(L$^1$)$_2$]$_3$\} was put
into a simulation cell of size 22$\times$22$\times$18 {\AA}, over
which the fast Fourier transform of the charge density has been
done with the cutoff 260 Ry. This corresponds to
216$\times$216$\times$180 divisions along the simulation cell
edges, and was tested to be sufficient to obtain converged energy
differences between different magnetic configurations. A discrete
energy spectrum was broadened in the following figures with the
halfwidth parameter of 0.1 eV in order to get a continuous density
of states (DOS). The calculation can be converged to both the
ferrimagnetic configuration (the ground state) and the
antiferromagnetic configuration (with the spin of one of the outer
Fe atoms set parallel to the central one), which is 0.13 eV higher
in energy.

\section[#4]{Electronic structure and charge state of F\lowercase{e}}
\label{sec:results}

\begin{figure}[b]
\centerline{\includegraphics[width=0.5\textwidth,clip=true]{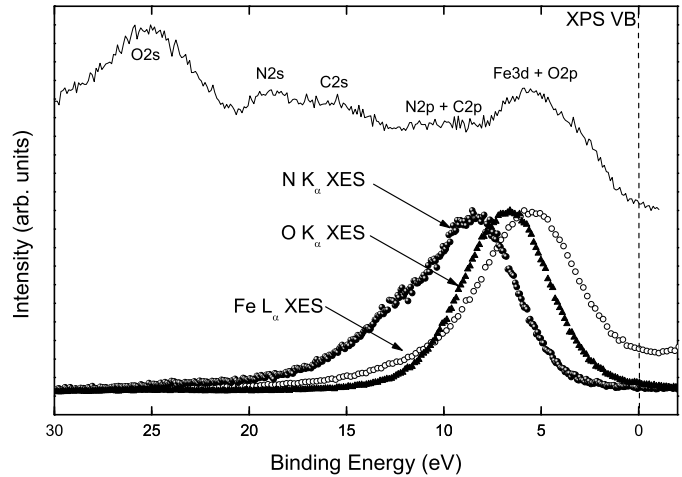}}
\caption{\label{fig:VB+XES}
Valence-band X-ray photoelectron spectrum of the Fe-``ferric
star'' and X-ray emission spectra of O, N, and Fe.}
\end{figure}

We start with a discussion of electronic structure of the ``ferric
star'' by comparing the spectroscopic data concerning the valence
band and first-principles calculations.
The XPS spectrum of the valence band is shown in
Fig.~\ref{fig:VB+XES} along with the element-selective X--ray
emission spectra of Fe, O, and N. The emission spectra in
Fig.~\ref{fig:VB+XES} correspond to normal non-resonant
fluorescence \cite{BESSY:fluor}. Taking into account the binding
energies of the corresponding core levels from the XPS (710.76 eV
for Fe $2p_{3/2}$, 531.96 eV for O $1s$, 400.76 eV for N $1s$),
the emission spectra were brought to a common energy scale with
the valence-band XPS, and they are plotted in
Fig.~\ref{fig:VB+XES}. This helps to identify the origin of
pronounced features in the valence band. The peak at 6 eV binding
energy is mostly due to Fe $3d4s$ states, but the O $2p$ states
contribute in about the same region (from $\sim$7 eV to higher
binding energy). The N $2p$ states are located significantly
deeper, forming a broad band centered at about 9 eV binding energy.

\begin{figure}[t]
\centerline{\includegraphics[width=0.5\textwidth,clip=true]{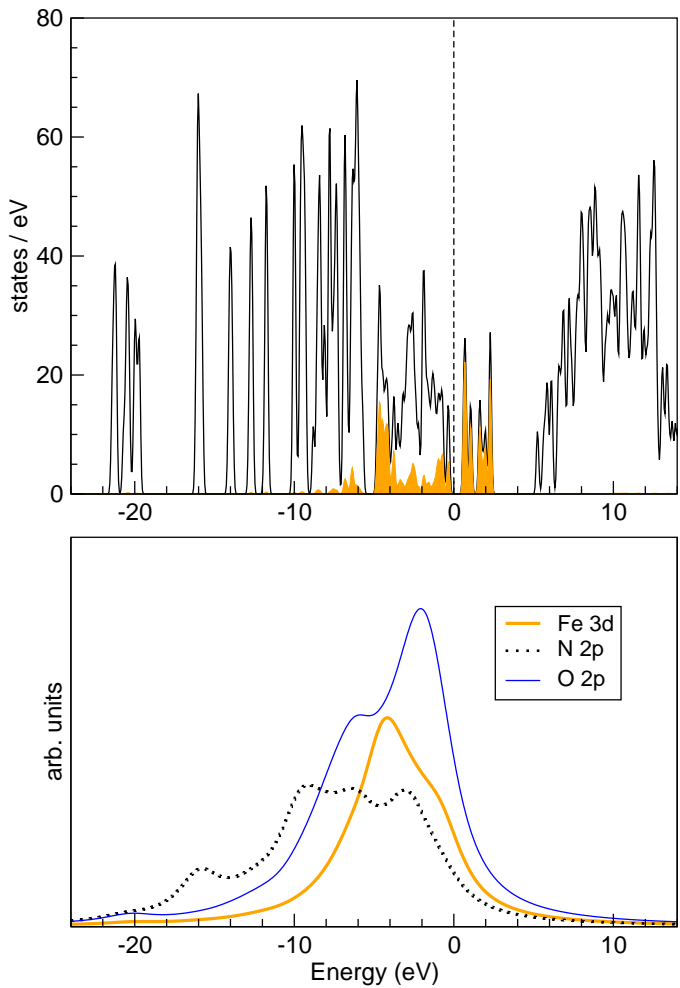}}
\caption{\label{fig:PDOS}
Top panel: calculated total density of
states of the Fe-``ferric star'' and local Fe$3d$ density of
states (shaded area). Bottom panel: local densities of occupied
O$2p$, Fe$3d$ and N$2p$ states, additionally convoluted with a
Lorentzian function of 1.6 eV halfwidth. See text for details.}
\end{figure}

These experimental observations are consistent with the results of
first-principle calculations, which produce the DOS (summed over
both spin directions) as shown in Fig.~\ref{fig:PDOS}. The top
panel depicts the total DOS (summed over both spin directions) in
the ground state. Zero energy separates occupied and vacant
states. One finds several bands, which are derived from the C, N,
and O $2s$ and $2p$ levels.
The lowest compact group of bands around $-$20 eV is almost
exclusively related to O $2s$ states. The region around the
``chemical potential'' ($E$=0) hosts Fe $3d$ states (shown shaded
in the figure), which are strongly spin-split (see the discussion
below) and strongly hybridize with the O $2p$ states. There is a
``band gap'' of $\approx$0.95 eV between the highest occupied and
the lowest unoccupied molecular orbitals, both being of mostly Fe
$3d$--O $2p$ character in the majority-spin channel (marked
spin-down in the following spin-resolved plot). The states with
the energies from $-$16 to $-$6 eV are all a mixture of different
orbitals forming covalent bonds and involving C $2s$,$2p$ and N
$2s$,$2p$ states.

\begin{figure}[b]
\centerline{\includegraphics[width=0.5\textwidth,clip=true]{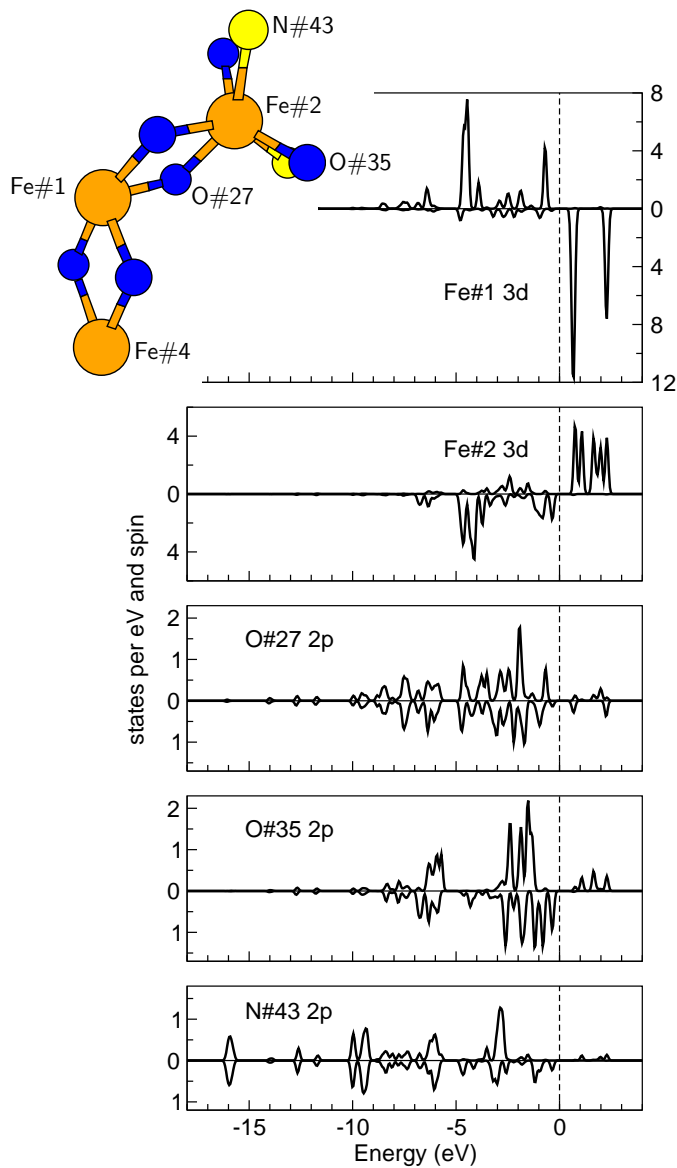}}
\caption{\label{fig:Fe_dos}
Spin-resolved local densities of states of
two inequivalent Fe sites and some their nearest neighbors. The
numbering of atoms in the relevant fragment of ``ferric star'' is
shown in the inset.}
\end{figure}

The bottom panel of Fig.~\ref{fig:PDOS} shows Fe $3d$, N $2p$ and
O $2p$ contributions, summed up over atoms of the same kind in
different positions in the molecule, and over both spin
directions; moreover an additional broadening was introduced to
yield a more straightforward comparison with the experimental XES
of Fig.~\ref{fig:VB+XES}. Indeed these three partial DOS roughly
span the states probed by the emission spectra, therefore the
comparison is complicated by the lack of structure in the
experimental spectra. One notices however a markedly lower energy
and much larger width of the N spectrum as compared to other two,
in both experiment and theory. The broadness of the N spectrum
indicates the participation of N $2p$ states in a number of
molecular orbitals, overlapping with C $2s$,$2p$ states throughout
the molecule.

Spin-resolved DOS for the atoms in which the spin splitting is
pronounced are shown in Fig.~\ref{fig:Fe_dos}. A corresponding
part of the molecule is shown in the inset, with the numbering of
atoms. The central atom Fe\#1, which has the bridge oxygen atoms
(\#27) as its only neighbors, carries a magnetic moment of 3.95
$\mu_{\mbox{\tiny B}}$. The outer Fe atoms (\#2) in the ground
state are magnetized oppositely to the central one, with magnetic
moments of $-$3.93 $\mu_{\mbox{\tiny B}}$, and have a different
environment, including two N atoms and two O\#35. The latter are
markedly magnetized, to $-$0.26 $\mu_{\mbox{\tiny B}}$. The bridge
oxygen atoms have a negligible net magnetic moment, but a marked
local spin density which changes sign along the path from Fe\#1 to
O\#27 to Fe\#2. The induced magnetization at the oxygen atoms adds
to the nominal spin moment associated with each Fe atom. The
``local'' moment of $\sim$4 $\mu_{\mbox{\tiny B}}$, which is
associated with Fe$3d$ states only, should be increased to $\sim$5
$\mu_{\mbox{\tiny B}}$ if one discusses a \emph{distributed}
magnetic moment, which in part resides over the ligands, and
follows the magnetization flips of its central Fe atom if they
occur. Hence, it is a ``well-behaved'' rigid moment in the sense
that the Heisenberg model, or another model dealing with well
defined spins, may be applied to it. These observations about the
local DOS, induced magnetizations of ligands, and a delocalized
but rigid spin moment associated with each Fe site are quite
similar to what has been reported earlier for ``ferric wheels'', a
chemically and structurally related class of molecular magnets
\cite{EMRS-Fewheel,Bedlewo-Fewheel}.

\begin{figure}[b]
\centerline{\includegraphics[width=0.5\textwidth,clip=true]{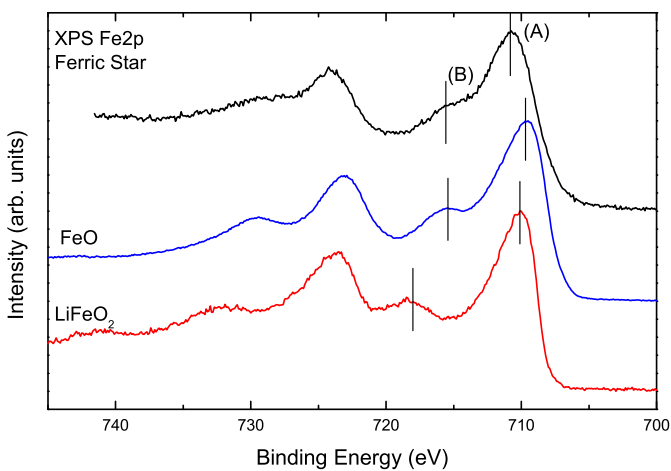}}
\caption{\label{fig:Fe2p}
Fe$2p$ photoelectron spectrum of the
Fe-``ferric star'' in comparison with those of FeO (definitely
Fe$^{2+}$) and LiFeO$_2$ (definitely Fe$^{3+}$). The features (A)
and (B) are discussed in the text.}
\end{figure}

It is essential to emphasize that bulk techniques (magnetization,
magnetic susceptibility) probe the effective spin value, however
without addressing its localization. The issue of localization --
whether the spin $S$=5 relates strictly to the Fe ion only or
extends over ligands -- is related to that of nominal valence.
Fe$^{3+}$ presumes the maximum-spin
$3d^5_{\uparrow}d^0_{\downarrow}$ configuration, whereas Fe$^{2+}$
corresponds to the local magnetic moment of 4 $\mu_{\mbox{\tiny
B}}$. Core-level spectroscopy, being an element-sensitive method,
may permit one to distinguish between these two cases.

\begin{figure}[t,h]
\centerline{\includegraphics[width=0.5\textwidth,clip=true]{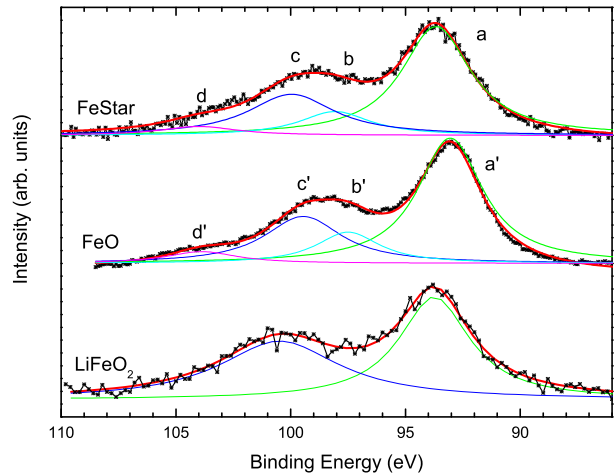}}
\caption{\label{fig:Fe3s}
Fe$3s$ photoelectron spectrum of the
Fe-``ferric star'' in comparison with those of FeO (definitely
Fe$^{2+}$) and LiFeO$_2$ (definitely Fe$^{3+}$). The features (a),
(b), (c) and (d) are discussed in the text.}
\end{figure}

Fig.~\ref{fig:Fe2p} depicts the $2p_{3/2}$ and $2p_{1/2}$
photoelectron spectra of the ``ferric star'' along with those of
two benchmark compounds, of well-known Fe$^{2+}$ and Fe$^{3+}$
valence: FeO, similar to what was earlier discussed in ref.
\cite{MayerB} and LiFeO$_2$ \cite{PRB56-4584}, respectively.
Characteristic ligand-to-metal charge-transfer effects are present
in the Fe $2p$ spectra of the Fe "ferric star", as well as in the
spectra of FeO and LiFeO$_2$, yet there are some dissimilarities
between these two reference systems. The relative positions and
widths of the peaks due to the final state with charge transfer
$2p^53d^{n+1}\underline{L}$ (A) and that without charge transfer
$2p^53d^nL$ (B) in the spectrum of ``ferric star'' is similar to
that of the FeO which indicates the Fe$^{2+}$ valence state.

The $3s$ core-level photoelectron spectra of the ``ferric star'',
again in comparison with those of FeO and LiFeO$_2$, are shown in
Fig.~\ref{fig:Fe3s}. The spectrum of the ``ferric star'' consists
of two well separated peaks ($a$) and ($c$), according to whether
the spin of the emitted $3s$ core electron is parallel or
antiparallel to that of the $3d$ shell. Each of these peaks has a
satellite denoted with ($b$) and ($d$), satellites which were
assigned by Sangaletti and Pamigiani \cite{JES98-287} in the case
of FeO to charge transfer excitations, with corresponding peaks
labelled $a'$, $b'$, $c'$ and $d'$.

In LiFeO$_2$ the charge transfer is
not prominent, therefore the fitting was done using two peaks
only. The splitting is proportional to the number of unpaired $d$
electrons and hence to total spin $S$; the predicted relative
intensities are given by the relation $S/(S+1)$
\cite{PRB71-085102}. The value of the Fe $3s$ splitting in the
case of the ``ferric star'' is about 5.10 eV, to be compared to
5.5 eV in FeO and 6.5 eV in LiFeO$_2$. We conclude that a nominal
valence Fe$^{2+}$ is more plausible in the ``ferric star''.

\section[#5]{Resonant X-ray emission in pure and C\lowercase{r}-doped
ferric stars} \label{sec:RIXS}

\begin{figure}[b]
\centerline{\includegraphics[width=0.5\textwidth,clip=true]{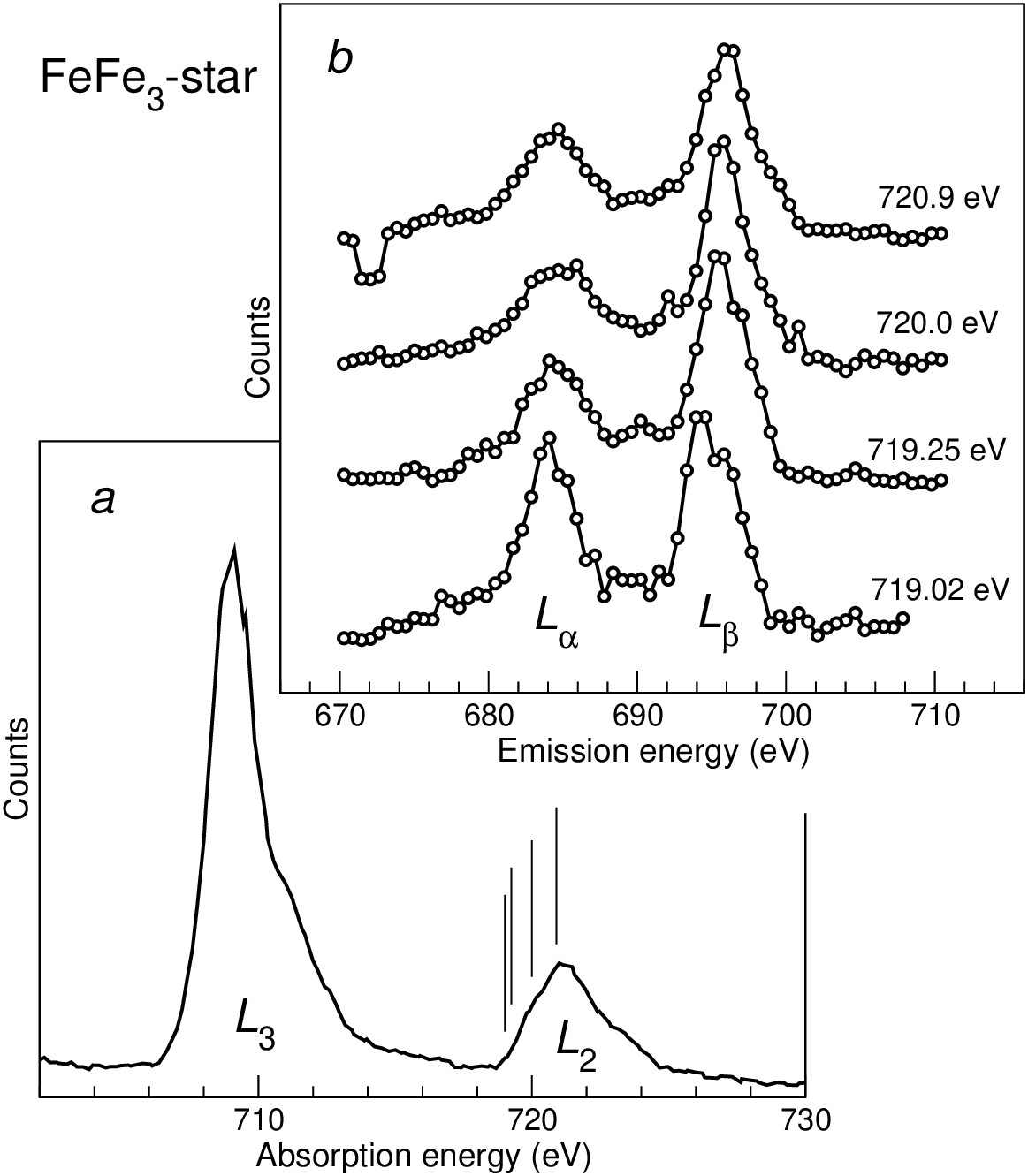}}
\caption{\label{fig:Fe-RIXS}
Near-edge X-ray absorption Fe $L_{2,3}$ spectrum (a) and
resonant inelastic X-ray Fe $L_3$ emission spectra (b) of the
Fe-``ferric star'' for four selected excitation energies, which
are also shown in the bottom panel as vertical bars. See text for
details.}
\end{figure}

\begin{figure}[b]
\centerline{\includegraphics[width=0.5\textwidth,clip=true]{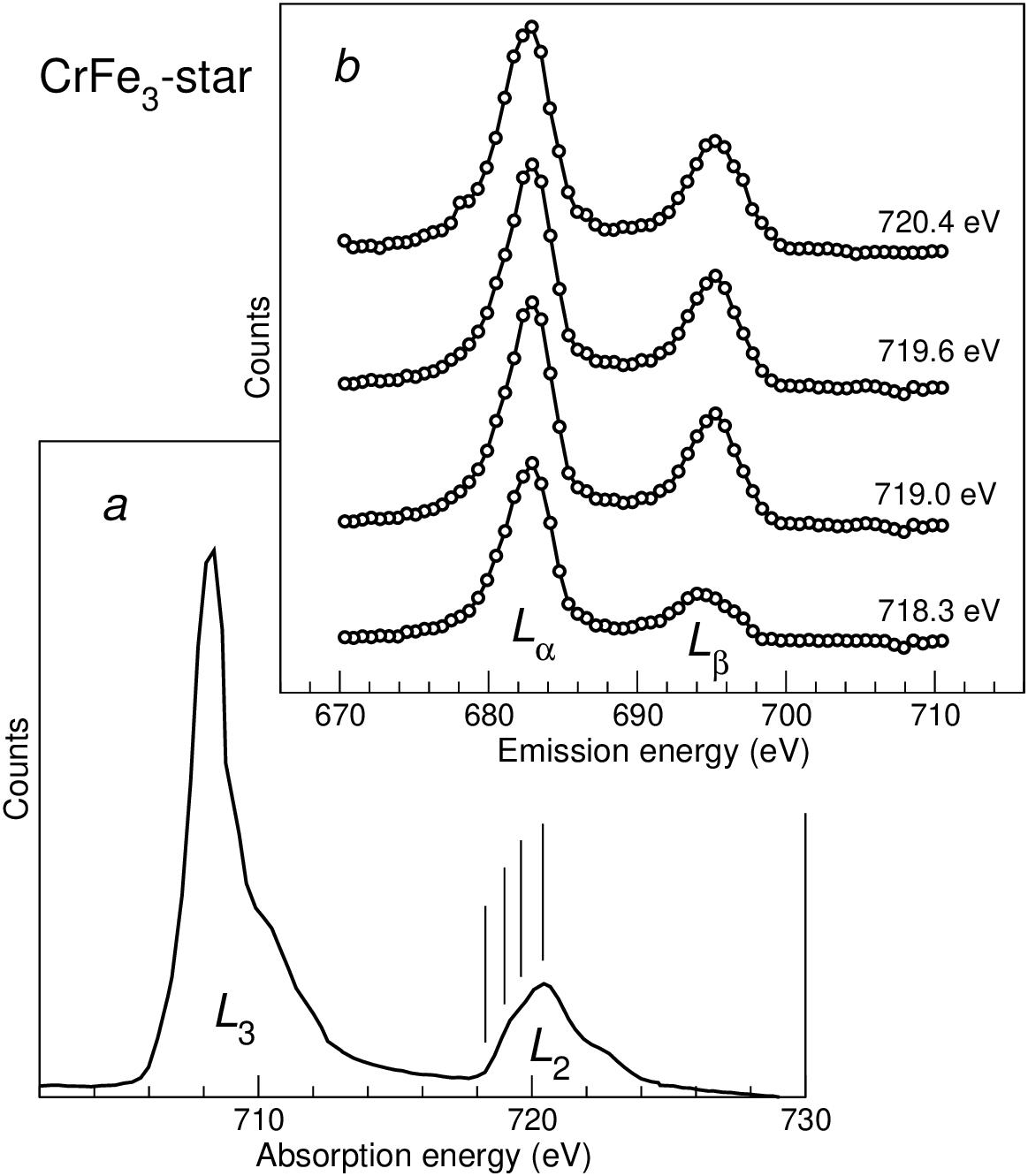}}
\caption{\label{fig:Cr-RIXS}
Near-edge X-ray absorption Fe $L_{2,3}$ spectrum (a) and
resonant inelastic X-ray Fe $L_3$ emission spectra (b) of the
Cr-``ferric star'' for four selected excitation energies, which
are also shown in the bottom panel as vertical bars. See text for
details.}
\end{figure}

We turn now to the comparative analysis of pure and Cr-doped
systems. The Cr substitution site is not unambiguously determined
by experiment, whereas recent first-principle calculations presume
that it must be the central one \cite{Cr-pos}. The valence-band
XPS and Fe $L$-emission spectra are very similar in both systems,
and Cr-related features cannot be resolved. However, certain
differences were found when analyzing resonant X-ray emission, see
Figs.~\ref{fig:Fe-RIXS} and \ref{fig:Cr-RIXS}. Panel (a) of
Fig.~\ref{fig:Fe-RIXS} depicts the NEXAFS spectrum of the ``ferric
star'' with $M$=Fe, covering the Fe $L_{2,3}$ edges. The peak at
about 709 eV corresponds to the ($L_3$) excitations from the Fe
$2p_{3/2}$ core state into the unoccupied $3d$ band, and the
second one at about 721 eV  corresponds to those ($L_2$) from the
Fe $2p_{1/2}$ state. The asymmetry seen in the first structure is
very similar to that observed for FeO, which has been explained in
terms of many overlapping multiplets \cite{PRB52-3143}. When
comparing this two-peaked structure with the XPS, we note that
here the $L_2$--$L_3$ splitting is 12.73 eV, i.e. less than the
spin-orbit splitting of the Fe$2p$ states determined by XPS (14
eV). The reason for the discrepancy is that the both the XPS and
absorption spectra consist of many multiplets, but these
multiplets are different for the ionic (XPS) and neutral
(absorption) final states, and therefore have different
splittings. Panel (b) of Fig.~\ref{fig:Fe-RIXS} shows the X-ray
emission spectra at the $L_3$ edge, obtained for four different
incident photon energies: 719.02, 719.25, 720, and 720.9 eV. Their
positions at the threshold and on the top of the $L_2$ edge are
indicated in the panel with the absorption spectrum.

In RIXS, a $2p{\rightarrow}$CB (CB: conduction band) excitation is
followed by a VB${\rightarrow}2p$ (VB: valence band) X-ray
emission; both events are normally treated as a single joint
process. As the excitation energy is gradually increased, on
reaching the $L_3$ absorption threshold the $L_{\alpha}$ emission
($3d{\rightarrow}2p_{3/2}$) becomes possible. As the incident
photon energy approaches the $L_2$ threshold, the $L_{\beta}$
emission ($3d{\rightarrow}2p_{1/2}$) appears as well. The
$L_{\alpha}$ emission persists, as an off-resonance scattering
process (creation and annihilation of a $2p_{3/2}$ hole), but also
possible by a Coster--Kronig process. In the latter, an initial
$2p_{1/2}$ hole is filled due to a radiationless
$2p_{3/2}{\rightarrow}2p_{1/2}$ transition, and the resulting
$2p_{3/2}$ hole is filled by $L_{\alpha}$ emission. These
different mechanisms affect the $L_{\alpha}/L_{\beta}$ intensity
ratio as a function of the excitation energy. We note that the
FeFe$_3$ and CrFe$_3$ stars, in spite of many similarities in
their Fe$L$ spectra, behave markedly differently in what regards
the $L_{\alpha}/L_{\beta}$ intensity ratio on passing through the
$L_2$ resonance excitation. In Fig.~\ref{fig:Fe-RIXS}(b) we
observe that at the maximum of the absorption resonance, the
$L_{\beta}$ intensity is higher than that of $L_{\alpha}$.
 We will see that this behavior is different from the
$L_{\alpha}/L_{\beta}$ intensity ratio which occurs in the case of
Cr substitution in the "ferric star".

Fig.~\ref{fig:Cr-RIXS}(a) depicts the NEXAFS at the Fe $L_{2,3}$
edge for the Cr-doped sample, and Fig.~\ref{fig:Cr-RIXS}(b) -- the
X-ray emission spectra, taken at the resonance energies 718.3,
719, 719.6 and 720.4 eV, from the onset to the maximum of the
$L_2$ absorption. The intensity of the $L_{\beta}$ emission first
grows and then remains constant, relative to that of $L_{\alpha}$.
The same behavior has been observed and discussed for the case of
FeO, in contrast to (non--magnetic) FeS$_2$, by Prince \emph{et
al.} \cite{PRB71-085102}. Another example of such behavior was
reported for Mn $L$ RIXS spectra in Heusler alloys
\cite{PRB63-235117}. The interpretation of results in these papers
related the $L_{\alpha}/L_{\beta}$ intensity to the high-spin
state of Fe or Mn. More specifically, the importance of having an
electronic structure with the majority-spin band occupied and the
minority-spin band empty was emphasized. In this case a sequence
of two processes, which may involve a Coster-Kronig transition as
an intermediate step -- an electron excitation into the CB and
filling of the core hole by an electron from the VB -- cannot
occur except with the inversion of spin, which is a
low-probability event. On the contrary, the electron excitation
into the CB and an immediate fluorescence transition, which is a
resonance process, remains allowed and explains the dominance of
the $L_{\beta}$ emission on $L_2$ excitation.

It is known that Coster--Kronig transition rates are also affected
by chemical environment. For example in the case of Te
\cite{Bahl}, changing the chemical environment drastically from
metallic to oxides to organometallic compounds produced a
variation of 23 {\%} in the ratio of the widths of the 3$p_{3/2}$
to 3$p_{1/2}$ peaks. In the present case, the chemical environment
is very similar so we expect small chemical effects. However the
$L_{\alpha}/L_{\beta}$ branching ratio changes by 2, which is a
large effect simillar to that in \cite{PRB71-085102} where the
changes was nearly a factor of 3. We therefore correlated the
large change observed to changes in magnetic properties.

\begin{figure}[t]
\centerline{\includegraphics[width=0.4\textwidth,clip=true]{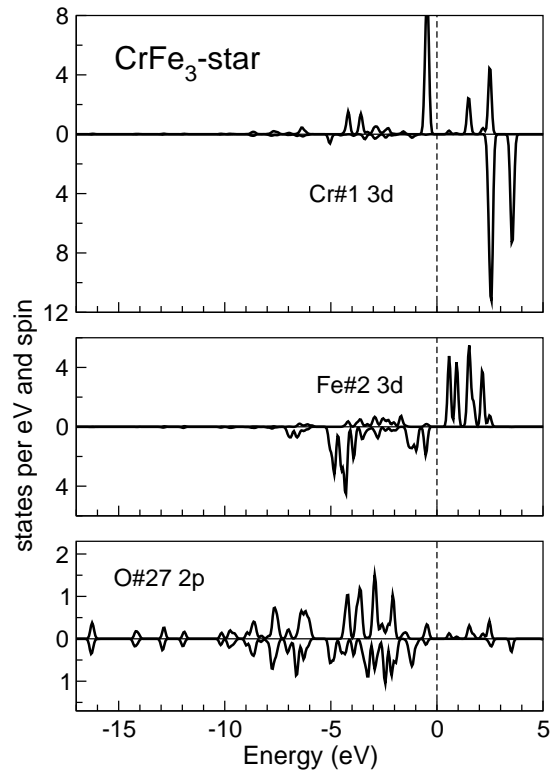}}
\caption{\label{fig:Cr_dos}
Spin-resolved local densities of states of Cr, Fe and the
bridging oxygen in the CrFe$_3$-star, to be compared to
Fig.~\ref{fig:Fe_dos}.}
\end{figure}

Fig.~\ref{fig:Fe_dos} makes clear that the local Fe DOS in the
``ferric star'' looks like that in a strong magnet, with the
majority-spin band fully occupied, and only a small contribution
(due to hybridization with O $2p$ states) present in the occupied
part of the minority-spin channel. Therefore the RIXS spectra in
the FeFe$_3$--star can be explained consistently with that in FeO
or in Heusler alloys. The situation with the CrFe$_3$--star, as is
seen from the calculated local DOS shown in Fig.~\ref{fig:Cr_dos},
is different: whereas the Fe local DOS (of the peripheral atoms,
compare to Fig.~\ref{fig:Fe_dos}) is identical to that in the
FeFe$_3$--star, the majority-spin states centered at the Cr site
are not fully occupied. Moreover the lowest unoccupied states
appear in the majority-spin channel. Therefore the
absorption-emission processes involving the Cr $3d$ states may
happen without the inversion of spin \cite{note_on_DOS}. However,
we admit that it is difficult to understand how this fact helps to
reduce an apparent blocking of the Coster-Kronig transitions
\emph{at the Fe site}. Some clue may be provided by considering
the bridging oxygen atoms (\#27 of Fig.~\ref{fig:Fe_dos}): whereas
in the FeFe$_3$--star the magnetization density changes its sign
exactly at the bridging O, in the CrFe$_3$--star this oxygen atom
is fully magnetized along with its Fe neighbor, and oppositely to
Cr. One can speculate that the matrix elements of the core-valence
transitions, which include differently composed molecular orbitals
in the valence band of FeFe$_3$--  and CrFe$_3$--stars, may affect
relative weights of $L_{\alpha}$ and $L_{\beta}$ emission. This is
however difficult to estimate without an explicit calculation of
the RIXS process.

We can, with relative certainty, eliminate the following other
hypothetical explanations of a different behavior in the
$L_{\alpha}/L_{\beta}$ intensity ratio of the CrFe$_3$--star. A
possibility of a low-spin state is ruled out by the calculation
result that in all trial spin-polarized calculations, the local
magnetic moments remained essentially fixed to
$\approx$3~$\mu_{\rm B}$ at the Cr site and $\approx$4~$\mu_{\rm
B}$ at Fe sites. There is no way to ``prepare'' the system with a
low-spin state of either Cr, or Fe \cite{stars_theo}. All
inversions of individual local moments from the ground state,
including the situation with the lowest total magnetic moment of 2
$\mu_{\rm B}$ in the
Cr$_{\uparrow}$Fe$_{\downarrow}$Fe$_{\downarrow}$Fe$_{\uparrow}$
star, preserve the ``strong magnetism'' at the Fe site and the
partial emptiness of the majority-spin states at the Cr site, with
strong similarities of their corresponding DOS to those shown in
Figs. \ref{fig:Fe_dos} and \ref{fig:Cr_dos}. Also the assumed Cr
substitution in the peripheral position (highly energetically
unfavorable, as was mentioned above) affected the magnetic moments
and local DOS only slightly. An additional inclusion of on-site
intraatomic Coulomb correlation beyond the conventional DFT
treatment may affect the energy placements of $3d$ levels in the
Cr and Fe DOS, but it can only enhance the ``strong magnetism'' at
the Fe site rather than ease it, in the presence of Cr dopant.

Therefore the understanding of $L_{\alpha}/L_{\beta}$ behavior in
the CrFe$_3$-star is incomplete on the basis of DFT calculations,
whereas an explanation of the behavior in FeFe$_3$--star seems
rather obvious, and consistent with the observed trend
\cite{PRB71-085102} for FeO, as compared to FeS$_2$. We can only
emphasize the observed difference in RIXS spectra of FeFe$_3$--
and CrFe$_3$--stars as an experimental finding, which might
stimulate additional research on these systems.

\section [#6]{Conclusions}
We have reported photoemission and photoabsorption spectra of
\{\emph{M}[Fe(L$^1$)$_2$]$_3$\}$\cdot$ 4CHCl$_3$ ($M$ = Fe, Cr) in
comparison with first-principles electronic structure
calculations. First we determined the charge state of Fe in these
``ferric stars'', in view of an existing controversy between the
expected nominal chemical valence (Fe III), from one side, and the
calculated value of the Fe local magnetic moment (4 $\mu_{\rm
B}$), that would imply a bivalent iron, from the other side. On
the basis of a comparison of the Fe $2p$ and $3s$ photoelectron
spectra with those of reference compounds, the charge state of Fe
in the FeFe$_3$--star was argued to be $2^+$. The valence-band
X-ray photoelectron spectrum and the X-ray Fe$L$, O and N$K$
emission spectra show little structure but are otherwise in
agreement with DFT calculations. In the FeFe$_3$-- and
CrFe$_3$--stars, the $L_{\alpha}/L_{\beta}$ intensity ratio in
resonant X-ray emission, varies according to the excitation energy
on going through the Fe $L_2$ threshold. The trend observed in the
FeFe$_3$--star is similar to that earlier reported by Yablonskikh
\emph{et al.} and Prince \emph{et al.}, and believed to be a
signature of ``high-spin'' structures, or more precisely of strong
magnetic systems, like e.g.  FeO or Heusler alloys. The
explanation of this behavior is consistent with the assumption
that the Coster--Kronig process probability is suppressed in
strong magnets. The DFT calculation does indeed support an
assumption of such strong magnetic behavior in the FeFe$_3$-star.
However, the CrFe$_3$-star does not show such a trend in the
$L_{\alpha}/L_{\beta}$ intensity, even though the local electronic
structure at the Fe site remains largely the same as in the
FeFe$_3$-star, and the loss of strong magnetic character appears
confined to the Cr center.

\section*{Acknowledgments}
AVP, AS, and SS gratefully acknowledge the financial support by
the Deutsche Forschungsgemeinschaft (SPP 1137 `Molecular
Magnets'), AT and KK acknowledge financial support of the PhD
program of Lower Saxony (Germany). Many enlightening discussions
with P.~M\"uller, R. Fink and S.~Bl\"ugel are appreciated. We
thank J.~Kortus for making available for comparison the
unpublished results of his \emph{ab initio} calculations on the
ferric stars, which have been done prior to those of the present
work. We thank the beamline scientists from the BACH beamline
(ELETTRA) for the overall excellent support which they provided.


\begin{thebibliography}{41}
\expandafter\ifx\csname natexlab\endcsname\relax\def\natexlab#1{#1}\fi
\expandafter\ifx\csname bibnamefont\endcsname\relax
  \def\bibnamefont#1{#1}\fi
\expandafter\ifx\csname bibfnamefont\endcsname\relax
  \def\bibfnamefont#1{#1}\fi
\expandafter\ifx\csname citenamefont\endcsname\relax
  \def\citenamefont#1{#1}\fi
\expandafter\ifx\csname url\endcsname\relax
  \def\url#1{\texttt{#1}}\fi
\expandafter\ifx\csname urlprefix\endcsname\relax\def\urlprefix{URL }\fi
\providecommand{\bibinfo}[2]{#2}
\providecommand{\eprint}[2][]{\url{#2}}

\bibitem[{\citenamefont{Verdaguer et~al.}(1999)\citenamefont{Verdaguer,
  Bleuzen, Train, Garde, Fabrizi {de}~Biani, and
  Desplanches}}]{PhilTransA357-1762}
\bibinfo{author}{\bibfnamefont{M.}~\bibnamefont{Verdaguer}},
  \bibinfo{author}{\bibfnamefont{A.}~\bibnamefont{Bleuzen}},
  \bibinfo{author}{\bibfnamefont{C.}~\bibnamefont{Train}},
  \bibinfo{author}{\bibfnamefont{R.}~\bibnamefont{Garde}},
  \bibinfo{author}{\bibfnamefont{F.}~\bibnamefont{Fabrizi {de}~Biani}},
  \bibnamefont{and}
  \bibinfo{author}{\bibfnamefont{C.}~\bibnamefont{Desplanches}},
  \bibinfo{journal}{Phil. Trans. R. Soc. Lond. A}
  \textbf{\bibinfo{volume}{357}}, \bibinfo{pages}{2959} (\bibinfo{year}{1999}).

\bibitem[{\citenamefont{Leuenberger and Loss}(2001)}]{Nat410-789}
\bibinfo{author}{\bibfnamefont{M.~N.} \bibnamefont{Leuenberger}}
  \bibnamefont{and} \bibinfo{author}{\bibfnamefont{D.}~\bibnamefont{Loss}},
  \bibinfo{journal}{Nature} \textbf{\bibinfo{volume}{410}},
  \bibinfo{pages}{789} (\bibinfo{year}{2001}).

\bibitem[{\citenamefont{Ohkoshi and Hashimoto}(2002)}]{ElChInt11-34}
\bibinfo{author}{\bibfnamefont{S.-i.} \bibnamefont{Ohkoshi}} \bibnamefont{and}
  \bibinfo{author}{\bibfnamefont{K.}~\bibnamefont{Hashimoto}},
  \bibinfo{journal}{The Electrochemical Society Interface}
  \textbf{\bibinfo{volume}{11}}, \bibinfo{pages}{34} (\bibinfo{year}{2002}).

\bibitem[{\citenamefont{Kahn}(1993)}]{Kahn-book}
\bibinfo{author}{\bibfnamefont{O.}~\bibnamefont{Kahn}},
  \emph{\bibinfo{title}{Molecular Magnetism}} (\bibinfo{publisher}{John Wiley
  \& Sons}, \bibinfo{address}{Singapore}, \bibinfo{year}{1993}), ISBN
  \bibinfo{isbn}{0-471-18838-7}.

\bibitem[{\citenamefont{Linert and Verdaguer}(2003)}]{Mol_Magnets}
\bibinfo{editor}{\bibfnamefont{W.}~\bibnamefont{Linert}} \bibnamefont{and}
  \bibinfo{editor}{\bibfnamefont{M.}~\bibnamefont{Verdaguer}}, eds.,
  \emph{\bibinfo{title}{Molecular Magnets}}
  (\bibinfo{publisher}{Springer-Verlag}, \bibinfo{address}{Wien},
  \bibinfo{year}{2003}), ISBN \bibinfo{isbn}{ISBN: 3-211-83891-0},
  \bibinfo{note}{special Edition of Monatshefte f{\"u}r Chemie/Chemical
  Monthly, Vol. 134, No. 2}.

\bibitem[{\citenamefont{Miller and Epstein}(2000)}]{MRSB25-21}
\bibinfo{author}{\bibfnamefont{J.~S.} \bibnamefont{Miller}} \bibnamefont{and}
  \bibinfo{author}{\bibfnamefont{A.~J.} \bibnamefont{Epstein}},
  \bibinfo{journal}{MRS Bulletin} \textbf{\bibinfo{volume}{25}},
  \bibinfo{pages}{21} (\bibinfo{year}{2000}).

\bibitem[{\citenamefont{Blundell and Pratt}(2004)}]{JPCM16-R771}
\bibinfo{author}{\bibfnamefont{S.~J.} \bibnamefont{Blundell}} \bibnamefont{and}
  \bibinfo{author}{\bibfnamefont{F.~L.} \bibnamefont{Pratt}},
  \bibinfo{journal}{J.~Phys.:~Condens.~Matter} \textbf{\bibinfo{volume}{16}},
  \bibinfo{pages}{R771} (\bibinfo{year}{2004}).

\bibitem[{\citenamefont{Saalfrank et~al.}(2001)\citenamefont{Saalfrank, Bernt,
  Chowdhry, Hampel, and Vaughan}}]{ChEurJ7-2765}
\bibinfo{author}{\bibfnamefont{R.~W.} \bibnamefont{Saalfrank}},
  \bibinfo{author}{\bibfnamefont{I.}~\bibnamefont{Bernt}},
  \bibinfo{author}{\bibfnamefont{M.~M.} \bibnamefont{Chowdhry}},
  \bibinfo{author}{\bibfnamefont{F.}~\bibnamefont{Hampel}}, \bibnamefont{and}
  \bibinfo{author}{\bibfnamefont{G.~B.~M.} \bibnamefont{Vaughan}},
  \bibinfo{journal}{Chem. Eur. J.} \textbf{\bibinfo{volume}{7}},
  \bibinfo{pages}{2765} (\bibinfo{year}{2001}).

\bibitem[{\citenamefont{Barra et~al.}(1999)\citenamefont{Barra, Caneschi,
  Cornia, Fabrizi {de}~Biani, Gatteschi, Sangregorio, Sessoli, and
  Sorace}}]{JACS121-5302}
\bibinfo{author}{\bibfnamefont{A.~L.} \bibnamefont{Barra}},
  \bibinfo{author}{\bibfnamefont{A.}~\bibnamefont{Caneschi}},
  \bibinfo{author}{\bibfnamefont{A.}~\bibnamefont{Cornia}},
  \bibinfo{author}{\bibfnamefont{F.}~\bibnamefont{Fabrizi {de}~Biani}},
  \bibinfo{author}{\bibfnamefont{D.}~\bibnamefont{Gatteschi}},
  \bibinfo{author}{\bibfnamefont{C.}~\bibnamefont{Sangregorio}},
  \bibinfo{author}{\bibfnamefont{R.}~\bibnamefont{Sessoli}}, \bibnamefont{and}
  \bibinfo{author}{\bibfnamefont{L.}~\bibnamefont{Sorace}},
  \bibinfo{journal}{J. Am. Chem. Soc.} \textbf{\bibinfo{volume}{121}},
  \bibinfo{pages}{5302} (\bibinfo{year}{1999}).

\bibitem[{\citenamefont{Cornia et~al.}(2004)\citenamefont{Cornia, Fabretti,
  Garrisi, Mortalo, Bonacchi, Gatteschi, Sessoli, Sorace, W.Wernsdorfer, and
  Barra}}]{AngewChem2004-43}
\bibinfo{author}{\bibfnamefont{A.}~\bibnamefont{Cornia}},
  \bibinfo{author}{\bibfnamefont{A.~C.} \bibnamefont{Fabretti}},
  \bibinfo{author}{\bibfnamefont{P.}~\bibnamefont{Garrisi}},
  \bibinfo{author}{\bibfnamefont{C.}~\bibnamefont{Mortalo}},
  \bibinfo{author}{\bibfnamefont{D.}~\bibnamefont{Bonacchi}},
  \bibinfo{author}{\bibfnamefont{D.}~\bibnamefont{Gatteschi}},
  \bibinfo{author}{\bibfnamefont{R.}~\bibnamefont{Sessoli}},
  \bibinfo{author}{\bibfnamefont{L.}~\bibnamefont{Sorace}},
  \bibinfo{author}{\bibnamefont{W.Wernsdorfer}}, \bibnamefont{and}
  \bibinfo{author}{\bibfnamefont{A.~L.} \bibnamefont{Barra}},
  \bibinfo{journal}{Angewandte Chemie -- International Edition}
  \textbf{\bibinfo{volume}{43}}, \bibinfo{pages}{1136} (\bibinfo{year}{2004}).

\bibitem[{\citenamefont{Saalfrank et~al.}(1997)\citenamefont{Saalfrank, Bernt,
  Uller, and Hampel}}]{AnChIE36-2482}
\bibinfo{author}{\bibfnamefont{R.~W.} \bibnamefont{Saalfrank}},
  \bibinfo{author}{\bibfnamefont{I.}~\bibnamefont{Bernt}},
  \bibinfo{author}{\bibfnamefont{E.}~\bibnamefont{Uller}}, \bibnamefont{and}
  \bibinfo{author}{\bibfnamefont{F.}~\bibnamefont{Hampel}},
  \bibinfo{journal}{Angewandte Chemie -- International Edition}
  \textbf{\bibinfo{volume}{36}}, \bibinfo{pages}{2482} (\bibinfo{year}{1997}).

\bibitem[{\citenamefont{Postnikov et~al.}(2004)\citenamefont{Postnikov,
  Chiuzb{\u{a}}ian, Neumann, and Bl{\"u}gel}}]{EMRS-Fewheel}
\bibinfo{author}{\bibfnamefont{A.~V.} \bibnamefont{Postnikov}},
  \bibinfo{author}{\bibfnamefont{S.~G.} \bibnamefont{Chiuzb{\u{a}}ian}},
  \bibinfo{author}{\bibfnamefont{M.}~\bibnamefont{Neumann}}, \bibnamefont{and}
  \bibinfo{author}{\bibfnamefont{S.}~\bibnamefont{Bl{\"u}gel}},
  \bibinfo{journal}{J. Phys. Chem. Solids} \textbf{\bibinfo{volume}{65}},
  \bibinfo{pages}{813} (\bibinfo{year}{2004}).

\bibitem[{\citenamefont{Postnikov et~al.}(2003)\citenamefont{Postnikov, Kortus,
  and Bl{\"u}gel}}]{Bedlewo-Fewheel}
\bibinfo{author}{\bibfnamefont{A.~V.} \bibnamefont{Postnikov}},
  \bibinfo{author}{\bibfnamefont{J.}~\bibnamefont{Kortus}}, \bibnamefont{and}
  \bibinfo{author}{\bibfnamefont{S.}~\bibnamefont{Bl{\"u}gel}},
  \bibinfo{journal}{Molecular Physics Reports} \textbf{\bibinfo{volume}{38}},
  \bibinfo{pages}{56} (\bibinfo{year}{2003}).

\bibitem[{\citenamefont{Gatteschi et~al.}(2000)\citenamefont{Gatteschi,
  Sessoli, and Cornia}}]{ChemComm2000-725}
\bibinfo{author}{\bibfnamefont{D.}~\bibnamefont{Gatteschi}},
  \bibinfo{author}{\bibfnamefont{R.}~\bibnamefont{Sessoli}}, \bibnamefont{and}
  \bibinfo{author}{\bibfnamefont{A.}~\bibnamefont{Cornia}},
  \bibinfo{journal}{Chem. Commun.} p. \bibinfo{pages}{725}
  (\bibinfo{year}{2000}).

\bibitem[{Kor()}]{Kortus_aniso}
\bibinfo{note}{J. Kortus (unpublished)}.

\bibitem[{\citenamefont{Koch et~al.}()\citenamefont{Koch, Schromm, Rupp, and
  M\"uller}}]{Mueller_aniso}
\bibinfo{author}{\bibfnamefont{R.}~\bibnamefont{Koch}},
  \bibinfo{author}{\bibfnamefont{S.}~\bibnamefont{Schromm}},
  \bibinfo{author}{\bibfnamefont{H.}~\bibnamefont{Rupp}}, \bibnamefont{and}
  \bibinfo{author}{\bibfnamefont{P.}~\bibnamefont{M\"uller}},
  \bibinfo{note}{private communication}.

\bibitem[{\citenamefont{Postnikov and Bl\"ugel}()}]{stars_theo}
\bibinfo{author}{\bibfnamefont{A.~V.} \bibnamefont{Postnikov}}
  \bibnamefont{and} \bibinfo{author}{\bibfnamefont{S.}~\bibnamefont{Bl\"ugel}},
  \bibinfo{note}{to be published}.

\bibitem[{\citenamefont{Zangrando et~al.}(2001)\citenamefont{Zangrando,
  Finazzi, Polucci, Comelli, Diciacco, Walker, Cocco, and Parmigiani}}]{BACH}
\bibinfo{author}{\bibfnamefont{M.}~\bibnamefont{Zangrando}},
  \bibinfo{author}{\bibfnamefont{M.}~\bibnamefont{Finazzi}},
  \bibinfo{author}{\bibfnamefont{G.}~\bibnamefont{Polucci}},
  \bibinfo{author}{\bibfnamefont{G.}~\bibnamefont{Comelli}},
  \bibinfo{author}{\bibfnamefont{B.}~\bibnamefont{Diciacco}},
  \bibinfo{author}{\bibfnamefont{R.~P.} \bibnamefont{Walker}},
  \bibinfo{author}{\bibfnamefont{D.}~\bibnamefont{Cocco}}, \bibnamefont{and}
  \bibinfo{author}{\bibfnamefont{F.}~\bibnamefont{Parmigiani}},
  \bibinfo{journal}{Rev. Sci. Instrum.} \textbf{\bibinfo{volume}{72}},
  \bibinfo{pages}{1313} (\bibinfo{year}{2001}).

\bibitem[{\citenamefont{Cocco et~al.}(2001)\citenamefont{Cocco, Matteucci,
  Prince, and Zangrando}}]{COMIXS}
\bibinfo{author}{\bibfnamefont{D.}~\bibnamefont{Cocco}},
  \bibinfo{author}{\bibfnamefont{M.}~\bibnamefont{Matteucci}},
  \bibinfo{author}{\bibfnamefont{K.}~\bibnamefont{Prince}}, \bibnamefont{and}
  \bibinfo{author}{\bibfnamefont{M.}~\bibnamefont{Zangrando}},
  \bibinfo{journal}{Proceedings SPIE} \textbf{\bibinfo{volume}{4506}}
  (\bibinfo{year}{2001}).

\bibitem[{\citenamefont{Beamson and Briggs}(1992)}]{Beamson}
\bibinfo{editor}{\bibfnamefont{G.}~\bibnamefont{Beamson}} \bibnamefont{and}
  \bibinfo{editor}{\bibfnamefont{D.}~\bibnamefont{Briggs}}, eds.,
  \emph{\bibinfo{title}{High Resolution {XPS} of Organic Polymers: The Scienta
  {ESCA300} Database}} (\bibinfo{publisher}{John Wiley \& Sons, Chichester},
  \bibinfo{year}{1992}).

\bibitem[{\citenamefont{Doniach and {\^S}unji{\'c}}(1970)}]{DS_func}
\bibinfo{author}{\bibfnamefont{S.}~\bibnamefont{Doniach}} \bibnamefont{and}
  \bibinfo{author}{\bibfnamefont{M.}~\bibnamefont{{\^S}unji{\'c}}},
  \bibinfo{journal}{Journal of Physics C} \textbf{\bibinfo{volume}{3}},
  \bibinfo{pages}{285} (\bibinfo{year}{1970}).

\bibitem[{\citenamefont{Tougaard}(1988)}]{Toug_func}
\bibinfo{author}{\bibfnamefont{S.}~\bibnamefont{Tougaard}},
  \bibinfo{journal}{Surface and Interface Analysis}
  \textbf{\bibinfo{volume}{11}}, \bibinfo{pages}{453} (\bibinfo{year}{1988}).

\bibitem[{\citenamefont{Ordej{\'o}n et~al.}(1996)\citenamefont{Ordej{\'o}n,
  Artacho, and Soler}}]{PRB53-10441}
\bibinfo{author}{\bibfnamefont{P.}~\bibnamefont{Ordej{\'o}n}},
  \bibinfo{author}{\bibfnamefont{E.}~\bibnamefont{Artacho}}, \bibnamefont{and}
  \bibinfo{author}{\bibfnamefont{J.~M.} \bibnamefont{Soler}},
  \bibinfo{journal}{Phys.~Rev.~B} \textbf{\bibinfo{volume}{53}},
  \bibinfo{pages}{R10441} (\bibinfo{year}{1996}).

\bibitem[{\citenamefont{Soler et~al.}(2002)\citenamefont{Soler, Artacho, Gale,
  Garc{\'{\i}}a, Junquera, Ordej{\'o}n, and S{\'a}nchez-Portal}}]{JPCM14-2745}
\bibinfo{author}{\bibfnamefont{J.~M.} \bibnamefont{Soler}},
  \bibinfo{author}{\bibfnamefont{E.}~\bibnamefont{Artacho}},
  \bibinfo{author}{\bibfnamefont{J.~D.} \bibnamefont{Gale}},
  \bibinfo{author}{\bibfnamefont{A.}~\bibnamefont{Garc{\'{\i}}a}},
  \bibinfo{author}{\bibfnamefont{J.}~\bibnamefont{Junquera}},
  \bibinfo{author}{\bibfnamefont{P.}~\bibnamefont{Ordej{\'o}n}},
  \bibnamefont{and}
  \bibinfo{author}{\bibfnamefont{D.}~\bibnamefont{S{\'a}nchez-Portal}},
  \bibinfo{journal}{J.~Phys.:~Condens.~Matter} \textbf{\bibinfo{volume}{14}},
  \bibinfo{pages}{2745} (\bibinfo{year}{2002}).

\bibitem[{sie()}]{siesta}
\urlprefix\url{http://www.uam.es/siesta}.

\bibitem[{\citenamefont{S{\'a}nchez-Portal
  et~al.}(1996)\citenamefont{S{\'a}nchez-Portal, Artacho, and
  Soler}}]{JPCM8-3859}
\bibinfo{author}{\bibfnamefont{D.}~\bibnamefont{S{\'a}nchez-Portal}},
  \bibinfo{author}{\bibfnamefont{E.}~\bibnamefont{Artacho}}, \bibnamefont{and}
  \bibinfo{author}{\bibfnamefont{J.~M.} \bibnamefont{Soler}},
  \bibinfo{journal}{J.~Phys.:~Condens.~Matter} \textbf{\bibinfo{volume}{8}},
  \bibinfo{pages}{3859} (\bibinfo{year}{1996}).

\bibitem[{\citenamefont{Junquera et~al.}(2001)\citenamefont{Junquera, Paz,
  S{\'a}nchez-Portal, and Artacho}}]{PRB64-235111}
\bibinfo{author}{\bibfnamefont{J.}~\bibnamefont{Junquera}},
  \bibinfo{author}{\bibfnamefont{{\'O}.}~\bibnamefont{Paz}},
  \bibinfo{author}{\bibfnamefont{D.}~\bibnamefont{S{\'a}nchez-Portal}},
  \bibnamefont{and} \bibinfo{author}{\bibfnamefont{E.}~\bibnamefont{Artacho}},
  \bibinfo{journal}{Phys.~Rev.~B} \textbf{\bibinfo{volume}{64}},
  \bibinfo{pages}{235111} (\bibinfo{year}{2001}).

\bibitem[{\citenamefont{Perdew et~al.}(1996)\citenamefont{Perdew, Burke, and
  Ernzerhof}}]{PRL77-3865}
\bibinfo{author}{\bibfnamefont{J.~P.} \bibnamefont{Perdew}},
  \bibinfo{author}{\bibfnamefont{K.}~\bibnamefont{Burke}}, \bibnamefont{and}
  \bibinfo{author}{\bibfnamefont{M.}~\bibnamefont{Ernzerhof}},
  \bibinfo{journal}{Phys.~Rev.~Lett.} \textbf{\bibinfo{volume}{77}},
  \bibinfo{pages}{3865} (\bibinfo{year}{1996}). 

\bibitem[{\citenamefont{Szargan et~al.}()\citenamefont{Szargan, Hallmeier,
  Hesse, Kopczynski, Krasnikov, Zhang, Chass\'e, Fuchs, Heske, and
  Umbach}}]{BESSY:fluor}
\bibinfo{author}{\bibfnamefont{R.}~\bibnamefont{Szargan}},
  \bibinfo{author}{\bibfnamefont{K.-H.} \bibnamefont{Hallmeier}},
  \bibinfo{author}{\bibfnamefont{R.}~\bibnamefont{Hesse}},
  \bibinfo{author}{\bibfnamefont{A.}~\bibnamefont{Kopczynski}},
  \bibinfo{author}{\bibfnamefont{S.~A.} \bibnamefont{Krasnikov}},
  \bibinfo{author}{\bibfnamefont{L.}~\bibnamefont{Zhang}},
  \bibinfo{author}{\bibfnamefont{T.}~\bibnamefont{Chass\'e}},
  \bibinfo{author}{\bibfnamefont{O.}~\bibnamefont{Fuchs}},
  \bibinfo{author}{\bibfnamefont{C.}~\bibnamefont{Heske}}, \bibnamefont{and}
  \bibinfo{author}{\bibfnamefont{E.}~\bibnamefont{Umbach}},
  \emph{\bibinfo{title}{X-ray fluorescence spectroscopy at {U41--PGM} by means
  of {ROSA}--present status and results}}, \bibinfo{note}{BESSY Annual Report
  2001}.

\bibitem[{\citenamefont{Mayer et~al.}(1996)\citenamefont{Mayer, Uhlenbrock, and
  Neumann}}]{MayerB}
\bibinfo{author}{\bibfnamefont{B.}~\bibnamefont{Mayer}},
  \bibinfo{author}{\bibfnamefont{S.}~\bibnamefont{Uhlenbrock}},
  \bibnamefont{and} \bibinfo{author}{\bibfnamefont{M.}~\bibnamefont{Neumann}},
  \bibinfo{journal}{J. Electron. Spectrosc. Relat. Phenom}
  \textbf{\bibinfo{volume}{81}}, \bibinfo{pages}{63} (\bibinfo{year}{1996}).

\bibitem[{\citenamefont{Galakhov et~al.}(1997)\citenamefont{Galakhov,
  Poteryaev, Kurmaev, Anisimov, Bartkowski, Neumann, Lu, Klein, and
  Zhao}}]{PRB56-4584}
\bibinfo{author}{\bibfnamefont{V.~R.} \bibnamefont{Galakhov}},
  \bibinfo{author}{\bibfnamefont{A.~I.} \bibnamefont{Poteryaev}},
  \bibinfo{author}{\bibfnamefont{E.~Z.} \bibnamefont{Kurmaev}},
  \bibinfo{author}{\bibfnamefont{V.~I.} \bibnamefont{Anisimov}},
  \bibinfo{author}{\bibfnamefont{S.}~\bibnamefont{Bartkowski}},
  \bibinfo{author}{\bibfnamefont{M.}~\bibnamefont{Neumann}},
  \bibinfo{author}{\bibfnamefont{Z.~W.} \bibnamefont{Lu}},
  \bibinfo{author}{\bibfnamefont{B.~M.} \bibnamefont{Klein}}, \bibnamefont{and}
  \bibinfo{author}{\bibfnamefont{T.-R.} \bibnamefont{Zhao}},
  \bibinfo{journal}{Phys.~Rev.~B} \textbf{\bibinfo{volume}{56}},
  \bibinfo{pages}{4584} (\bibinfo{year}{1997}).

\bibitem[{\citenamefont{Parmigiani and Sangaletti}(1999)}]{JES98-287}
\bibinfo{author}{\bibfnamefont{F.}~\bibnamefont{Parmigiani}} \bibnamefont{and}
  \bibinfo{author}{\bibfnamefont{L.}~\bibnamefont{Sangaletti}},
  \bibinfo{journal}{J. Electron Spectrosc. Relat. Phenom.}
  \textbf{\bibinfo{volume}{98--99}}, \bibinfo{pages}{287}
  (\bibinfo{year}{1999}).

\bibitem[{\citenamefont{Prince et~al.}(2005)\citenamefont{Prince, Matteucci,
  Kuepper, Chiuzb{\^a}ian, Bartkowski, and Neumann}}]{PRB71-085102}
\bibinfo{author}{\bibfnamefont{K.~C.} \bibnamefont{Prince}},
  \bibinfo{author}{\bibfnamefont{M.}~\bibnamefont{Matteucci}},
  \bibinfo{author}{\bibfnamefont{K.}~\bibnamefont{Kuepper}},
  \bibinfo{author}{\bibfnamefont{S.~G.} \bibnamefont{Chiuzb{\^a}ian}},
  \bibinfo{author}{\bibfnamefont{S.}~\bibnamefont{Bartkowski}},
  \bibnamefont{and} \bibinfo{author}{\bibfnamefont{M.}~\bibnamefont{Neumann}},
  \bibinfo{journal}{Phys.~Rev.~B} \textbf{\bibinfo{volume}{71}},
  \bibinfo{pages}{085102} (\bibinfo{year}{2005}).

\bibitem[{Cr-()}]{Cr-pos}
\bibinfo{note}{According to our recent calculations, the central position of
  substitutional Cr is by 1 eV/molecule more favorable than the peripherical
  one.}

\bibitem[{\citenamefont{Crocombette et~al.}(1995)\citenamefont{Crocombette,
  Pollak, Jollet, Thromat, and Gautier-Soyer}}]{PRB52-3143}
\bibinfo{author}{\bibfnamefont{J.~P.} \bibnamefont{Crocombette}},
  \bibinfo{author}{\bibfnamefont{M.}~\bibnamefont{Pollak}},
  \bibinfo{author}{\bibfnamefont{F.}~\bibnamefont{Jollet}},
  \bibinfo{author}{\bibfnamefont{N.}~\bibnamefont{Thromat}}, \bibnamefont{and}
  \bibinfo{author}{\bibfnamefont{M.}~\bibnamefont{Gautier-Soyer}},
  \bibinfo{journal}{Phys.~Rev.~B} \textbf{\bibinfo{volume}{52}},
  \bibinfo{pages}{3143} (\bibinfo{year}{1995}).

\bibitem[{\citenamefont{Yablonskikh et~al.}(2001)\citenamefont{Yablonskikh,
  Yarmoshenko, Grebennikov, Kurmaev, Butorin, Duda, Nordgren, Plogmann, and
  Neumann}}]{PRB63-235117}
\bibinfo{author}{\bibfnamefont{M.~V.} \bibnamefont{Yablonskikh}},
  \bibinfo{author}{\bibfnamefont{Y.~M.} \bibnamefont{Yarmoshenko}},
  \bibinfo{author}{\bibfnamefont{V.~I.} \bibnamefont{Grebennikov}},
  \bibinfo{author}{\bibfnamefont{E.~Z.} \bibnamefont{Kurmaev}},
  \bibinfo{author}{\bibfnamefont{S.~M.} \bibnamefont{Butorin}},
  \bibinfo{author}{\bibfnamefont{L.-C.} \bibnamefont{Duda}},
  \bibinfo{author}{\bibfnamefont{J.}~\bibnamefont{Nordgren}},
  \bibinfo{author}{\bibfnamefont{S.}~\bibnamefont{Plogmann}}, \bibnamefont{and}
  \bibinfo{author}{\bibfnamefont{M.}~\bibnamefont{Neumann}},
  \bibinfo{journal}{Phys.~Rev.~B} \textbf{\bibinfo{volume}{63}},
  \bibinfo{pages}{235117} (\bibinfo{year}{2001}).

\bibitem[{\citenamefont{Bahl et~al.}(1977)\citenamefont{Bahl, Watson, and
  Irgolic}}]{Bahl}
\bibinfo{author}{\bibfnamefont{M.~K.} \bibnamefont{Bahl}},
  \bibinfo{author}{\bibfnamefont{R.~L.} \bibnamefont{Watson}},
  \bibnamefont{and} \bibinfo{author}{\bibfnamefont{K.~J.}
  \bibnamefont{Irgolic}}, \bibinfo{journal}{J. Phys. Chem.}
  \textbf{\bibinfo{volume}{66}}, \bibinfo{pages}{5526} (\bibinfo{year}{1977}).

\bibitem[{not()}]{note_on_DOS}
\bibinfo{note}{We emphasize that the occupation of Fe or Cr states in the
  FeCr$_3$ star remains qualitatively the same also for other possible
  configurations: when the Cr atom substitutes a peripheral site, or with
  different setting of magnetization directions over Fe, Cr atoms.}

\end{thebibliography}
\end{document}